\documentstyle[12pt]{article}

\input BoxedEPS.tex
\SetRokickiEPSFSpecial
\HideDisplacementBoxes

\font\sf=cmss10                    

\def\del{\partial}
\def\Dslash{\not{\hbox{\kern-4pt $D$}}}
\def\dslash{\not{\hbox{\kern-2pt $\del$}}}
\def\gslash{\not{\hbox{\kern-2pt $\gamma$}}}
\def\g5{\gamma_5}

\newcommand{\mathbold}[1]{\mbox{\boldmath $\bf#1$}}
\def\mJ{\mathbold{J}}
\def\malpha{\mathbold{\alpha}}

\def\mA{\mathbold{A}}
\def\mB{\mathbold{B}}
\def\mC{\mathbold{C}}
\def\ma{\mathbold{a}}
\def\mb{\mathbold{b}}
\def\mc{\mathbold{c}}

\def\momega{\mathbold{\omega}}
\def\mdelta{\mathbold{\delta}}
\def\mSigma{\mathbold{\Sigma}}
\def\mtheta{\mathbold{\theta}}
\def\mOmega{\mathbold{\Omega}}
\def\bbbz{{\sf Z\!\!\!Z}}
\def\sl2z{SL(2,\bbbz)}
\def\fracs#1#2{\textstyle\frac #1#2}
\def\g{\cal{g}}
\def\nt{\tilde{n}}

\begin{document}
\setlength{\parindent}{.60in}  

\setlength{\baselineskip}{18pt} 

\noindent

\setlength{\parskip}{1.35ex}
\setlength{\parindent}{0em}

\thispagestyle{empty}
{\flushright{\small MIT-CTP-2776\\hep-th/9809026\\}}

\vspace{.3in}
\begin{center}\Large {\bf Affine Lie Algebras, String 
                          Junctions \\ and 7-Branes}
\end{center} 

\vspace{.1in}
\begin{center}
{\large Oliver DeWolfe}

\vspace{.1in}
{ {\it Center for Theoretical Physics,\\
Laboratory for Nuclear Science,\\
Department of Physics\\
Massachusetts Institute of Technology\\
Cambridge, Massachusetts 02139, U.S.A.}}
\vspace{.2in}

E-mail: {\tt odewolfe@ctp.mit.edu}
\end{center}
\begin{center}September 1998\end{center}

\vspace{0.1in}
\begin{abstract}
We consider the realization of affine ADE Lie algebras as string
junctions on mutually non-local 7-branes in Type IIB string theory.
The existence of the affine algebra is signaled by the presence of the
imaginary root junction $\mdelta$, which is realized as a string
encircling the 7-brane configuration.  The level $k$ of an affine
representation partially constrains the asymptotic $(p,q)$ charges of
string junctions departing the configuration.  The junction
intersection form reproduces the full affine inner product, plus terms
in the asymptotic charges.
\end{abstract}

\newpage

\section{Introduction}

The compactification of F-Theory on elliptically fibered K3 manifolds
has proven to be a fruitful testing ground for non-Abelian gauge
symmetry enhancement.  The singular fibers of the K3 are organized
according to the Kodaira classification, which associates to each an
$ADE$ Lie algebra.  From the IIB point of view the situation is a
compactification on $S^2$, with 24 mutually non-local 7-branes. In this
perspective the singularities are coinciding 7-branes and the gauge
vectors are strings or string junctions with support on these branes.
How the Lie-algebraic properties of the junctions arise from the
intersection of the corresponding holomorphic curves in K3 has been
extensively investigated \cite{GZ,DZ}.

The Kodaira classification includes singularities corresponding to
$A$-, $D$- and $E$-type Lie algebras.  Since these exhaust the
singularities on elliptic K3, only 7-branes corresponding to these
algebras can collapse to a point.  However, it is natural to wonder
about other configurations of 7-branes.  The Lie-algebraic weight
vector associated to a string junction on a configuration of 7-branes
is well-defined regardless of whether those branes can coalesce.  If
the root system of an algebra arises on a set of branes, the
associated junctions must organize themselves into representations of
that algebra, regardless of whether the gauge bosons can be massless.
Hence one is led to wonder about the possibility of realizing other
Lie algebras on 7-brane configurations.

Affine Lie algebras have played a prominent role in string theory.
Because the worldsheet currents of the heterotic string define an
affine algebra, massive stringy excitations organize themselves
according to affine representations.  Generalized Kac-Moody algebras
have been used to investigate spaces of BPS states \cite{moore}.
Furthermore, it is now well-known that there exists a 6D non-critical
string carrying an $E_8$ current algebra, related to zero-size
instantons \cite{ncsgh,ncssw}.  This non-critical string is
tensionless at certain points in moduli space, leading to the
possibility of an infinite tower of massless states organized
according to an affine representation in the theory compactified to
five or four dimensions \cite{vafa,ganor, minahan}.

Since the F-theory compactification is dual to the heterotic string,
the obvious question to ask is whether affine algebras can arise on
7-branes.  We find that indeed they do.  Certain collections of
7-branes admit a junction $\mdelta$ which is a string loop of some
$(p,q)$ charge winding entirely around the configuration; this
situation was examined in a particular case by \cite{imamura}.  This
plays the role of the imaginary root, the distinguishing element of
the affine root system.  We find that the level $k$ of the affine
representation associated to a set of junctions constrains one linear
combination of the asymptotic charges $(p,q)$.  The junction
intersection form includes the affine inner product, inducing the
affine Cartan matrix on the junctions without asymptotic charge
(hereafter, ``uncharged'' junctions) and including the ``light-cone''
product of the level and grade.  Decoupling a single brane leaves
behind a familiar $ADE$ configuration.

Affine representations are all infinite-dimensional.  This is due to
the presence of arbitrarily large numbers of $\mdelta$ contributions.
Since the affine singularity cannot be realized in K3, however, all
but a finite number of junctions (corresponding to the associated
finite algebra) must remain massive.  The necessity of ordinary finite
Lie algebra representations organizing themselves into affine
representations in the presence of an additional brane gives
information about the representations that can arise on the
worldvolume of a probe D3-brane.  The 4D $E_n$ theories associated to
the non-critical string were investigated from the point of view of a
3-brane/7-brane system in \cite{DHIZ}.

Section 2 reviews basic properties of affine Lie algebras.  Section 3
details the way in which affine algebras arise on 7-branes.  Section 4
explores these ideas in the simplest example of $\widehat{su(2)}$ and
the more complicated case of $\widehat{E_8}$ as well.  Section 5
contains some concluding remarks.  We do not consider the issue of
which junction of an equivalence class is the BPS representative,
which has been addressed elsewhere \cite{ghz,hauer}.

\section{Review of Affine Lie Algebras}

In this section we briefly review the characteristics of affine Lie algebras
that are relevant to our investigation.  For the definitive mathematical
treatment one should consult \cite{kac}, while discussions
aimed at physicists can be found in \cite{phys}.

Lie algebras are determined entirely by their Cartan matrices by the
Chevalley-Serre construction. An $r \times r$ matrix
$A_{ij}$ defines a Lie algebra $g$ by directly specifying
the brackets of a subset of the generators, the $3r$ Chevalley
generators $\{H^i, E^\pm_i\}, \, i = 1 \ldots r$:
\begin{eqnarray}
\label{chevalley}
[H^i, H^j] = 0 \,, \quad [H^i, E^\pm_j] = \pm A_{ji} E^\pm_j \,, \quad 
[E^+_i, E^-_j] = \delta^{ij} H^j \,.
\end{eqnarray}
The remaining elements of the algebra are obtained from successive
commutators of the Chevalley generators, modulo the Serre relations:
\begin{eqnarray}
\label{serre}
( \mbox{ad} \, E^\pm_i)^{1-A_{ji}} E^j_\pm = 0 \,, \,  i \neq j \,.
\end{eqnarray}
A finite, semisimple Lie algebra is obtained when the matrix $A$ has integral
entries and satisfies certain restrictions:
\begin{eqnarray}
A_{ii} = 2 \,; \quad A_{ij} \leq 0 \,, \, i \neq j \,; \quad A_{ij} = 0
\leftrightarrow A_{ji} = 0 \,;  \quad
\mbox{det} \, A > 0 \,.
\end{eqnarray}
An affine Cartan matrix is obtained by allowing $\mbox{det} \, A = 0$,
while requiring that removing the row and column associated to any one
element leaves a semisimple Cartan matrix.  Hence the affine Lie
algebras are simple generalizations of finite Lie algebras from this
point of view.

The degeneracy of its Cartan matrix is responsible for all the affine algebra's
unusual properties.  In particular, the algebra and its root system are
infinite.  The degeneracy results in the existence of
a linear combination of roots that has vanishing inner product with all roots,
the ``imaginary'' root $\delta$:
\begin{eqnarray}
\delta = \alpha_0 + \sum_{i=1}^r c^i  \, \alpha_i \,,
\end{eqnarray}
where the Coxeter labels $c^i$ are the expansion coefficients of the
highest root $\theta$ of $g$ in the basis $\{ \alpha_i \}$, $\theta =
\sum_{i=1}^r c^i \, \alpha_i$.\footnote{We will be dealing exclusively
with simply laced algebras, and so will not distinguish between
Coxeter labels and dual Coxeter labels.}  The existence of $\delta$
satisfying $(\delta,\alpha_i) = 0$ is the distinguishing characteristic
of an affine root system.

Because the junction realizations of Lie algebras originate with the
Cartan matrix, we have emphasized understanding how the affine Lie
algebra is a simple generalization of a finite algebra.  We now
examine the construction of an affine Lie algebra as a centrally
extended loop algebra with derivation, which is the usual presentation
of an affine algebra and is most useful for understanding weight
vectors and their inner product.

A loop algebra is a map from the circle $S^1$ to a Lie algebra $g$.
If $\{ T^a \}$ are a basis for $g$, the loop algebra has basis
$\{ T^a_n \} \equiv \{ T^a \otimes z^n \}$.  This loop algebra can be given
a nontrivial central extension by adding the generator $K$, satisfying
$[K, T^a_n] =0$, so that the brackets are
\begin{eqnarray}
[T^a_n, T^b_m] = f^{abc} \, T^c_{n+m} + n \, \delta_{n+m,0} \, \delta^{ab} \, 
K \,,
\end{eqnarray}
where we have diagonalized the Killing form of $g$.  Additionally one
includes the derivation $D$ (sometimes called $L_0$) which measures
the grade $n$:
\begin{eqnarray}
[D, T^a_n] = n \, T^a_n \,, \quad [D,K] = 0 \,.
\end{eqnarray}
These generators and their brackets define the affine algebra $\hat{g}$.
Notice that the elements with $n=0$ generate a subalgebra
isomorphic to $g$:
\begin{eqnarray}
[T^a_0, T^b_0] = f^{abc} \, T^c_0 \,,
\end{eqnarray}
called the horizontal subalgebra of $\hat{g}$; since
$\hat{g}_{\mbox{hor}} \cong g$, we shall use them interchangeably. 

Using a Cartan-Weyl basis for $\hat{g}$, the algebra is generated by
$\{H^i_n, E^\alpha_n, K, D\}.$  The Cartan generators can be chosen
to be $\{ H^i_0, K, D \}$, and their eigenvalues will characterize 
a weight vector for any state:
\begin{eqnarray}
\lambda = (\overline\lambda, k, n) \,, \label{weight}
\end{eqnarray}
 where $\overline\lambda$ is a horizontal weight, $k$ is the eigenvalue
of $K$ and is called the level, and $n$ is the grade. 

The roots are the weights of the adjoint representation.  They are
$(\overline\alpha, 0, n)$ corresponding to the generators $\{E^\alpha_n\}$ 
and $(0,0,n)$ corresponding to the generators $\{H^i_n\}$; note that since
$K$ is central all roots have vanishing level.  Simple roots are chosen 
such that an arbitrary root can be expressed as a linear combination 
with coefficients of definite sign.  We choose the set
\begin{eqnarray}
\alpha_i = (\overline\alpha_i, 0, 0) \,, \quad \alpha_0 = (-\overline\theta, 0 , 1) \,,
\end{eqnarray}
where $\overline\alpha_i$ are the simple roots of $g$ and $\overline\theta$
is the highest root of $g$. Additionally, we define the imaginary root
$\delta$:
\begin{eqnarray}
\delta \equiv (0,0,1) \,;  \quad \delta = \alpha_0 + \theta =  \alpha_0 +
\sum_{i=1}^r c^i \, \alpha_i   \,,
\label{delta}
\end{eqnarray}
where $r$ is the rank of $g$.  Note this is the same $\delta$ we encountered
before.

As with simple Lie algebras, an inner product is induced on the space of
weights by means of the Killing form.  This inner product is interesting
for its ``light-cone'' structure on the level and grade of
weights:
\begin{eqnarray}
(\lambda_1, \lambda_2) = (\overline\lambda_1, \overline\lambda_2) + k_1 \, n_2
+ k_2 \, n_1 \,.  \label{lightcone}
\end{eqnarray}
The inner product of the simple roots gives the affine Cartan matrix.
From this point of view $\delta$ has vanishing inner product with all roots
precisely because they have $k=0$.

As a consequence of $a_i = (\lambda, \alpha_i)$ and $k = (\lambda, \delta)$
for some weight $\lambda$ and the definition (\ref{delta}), we have
\begin{eqnarray}
k &=& a_0 + \sum_{i=1}^r c^i \, a_i \,,
\end{eqnarray}
and thus the $\{ a_i \}, i=0 \ldots r$ determine $k$, or conversely
$k$ and $\overline\lambda$ determine $a_0$.

The adjoint is not a highest weight representation.  Highest weight
representations do exist, and are all infinite-dimensional.  They are
formed by beginning with a highest weight and subtracting simple roots
according to the values of the Dynkin labels as usual.  Due to the
degeneracy of the Cartan matrix, there is no end to the process, and
the grade $n$ never stops decreasing --- $\delta$ can always be
subtracted from some weight without leaving the representation.  For
the ``integrable'' highest weight representations, however, there will
be a finite (reducible) representation of the horizontal subalgebra at
each grade.  Thus the highest weight representations have a pyramid
structure.  The level $k$ is a constant for an entire
representation, as the simple roots have $k=0$.  The condition for
integrability is just that for the highest weight all Dynkin labels,
$a_0 \ldots a_r$, must be non-negative integers.  This implies that the
horizontal piece is a horizontal highest weight, and that the level is
bounded below
\begin{eqnarray}
k  \geq \sum_{i=1}^8 c^i \, a_i \geq 0 \,.
\end{eqnarray}
The grade of the highest weight vector is conventionally
normalized to $n_0 = 0$.

\section{Generalities of affine algebras on branes}

String junctions with support on configurations of mutually nonlocal
7-branes fill out representations of a Lie algebra determined by the
7-branes.  A given 7-brane is characterized by the NSNS and RR charges
$[p,q]$ of the string which can end on it.  Each 7-brane will
influence a junction $\mJ$ through a combination of two processes:
segments of $\mJ$ may have a string prong ending on the 7-brane, or
they may undergo a monodromy in crossing the 7-brane's branch cut.
Both situations transfer a certain amount of $(p,q)$ charge to the
relevant segment of the junction, and can be deformed into one another
via a Hanany-Witten effect \cite{hanany}.  Both are taken into account by the
invariant charge $Q^a(\mJ)$, the integer multiple of $[p_a,q_a]$
added to the junction $\mJ$ by the $a^{th}$ 7-brane in toto \cite{DZ}.
Since the $\{ Q^a \}$ are integral, the space of inequivalent
junctions is a lattice.  The Lie-algebraic weight vector $\vec\lambda$
associated to $\mJ$ is determined by the invariant charges.

An inner product exists on the space of junctions, arising from the
intersection of complex curves in M/F-Theory which reduce to the
junctions in IIB \cite{mns,iqbal}.  The junctions corresponding to
gauge bosons in the 8D worldvolume begin and end on the 7-branes and
so carry no asymptotic $(p,q)$ charge to infinity.  This constrains
the adjoint junctions to live in a sublattice of codimension two.  The
simple root states $\{ \malpha_i \}$ can be chosen to span this
sublattice, and their mutual intersection form produces (minus) the
Cartan matrix of the appropriate Lie algebra.  The Dynkin labels
characterizing the weight vector associated to a junction $\mJ$ are
given by $a_i = - (\mJ, \malpha_i)$.

Since the Lie algebras on 7-branes are generated by the Cartan
matrices, an affine Lie algebra should arise when the uncharged
junctions produce an affine Cartan matrix in their intersection form.
Thus despite the complications of the affine algebra, it will appear
when just a single brane is added appropriately to an $ADE$
configuration.

Once the simple roots for the affine algebra are identified, the
entire root system follows.  The roots for the horizontal part can be
found as they were in the finite case \cite{DZ}.  Due to the
degeneracy of the Cartan matrix, the imaginary root junction
\begin{eqnarray}
\mdelta = \malpha_0 + \sum_{i=1}^r c^i \, \malpha_i \,.
\end{eqnarray}
has vanishing intersection with all simple roots, and thus with all
roots, including itself.  Thus the junctions of self-intersection
$(-2)$ are precisely the horizontal roots $\malpha$ plus arbitrary
factors of $\mdelta$, $\{ \malpha + n \, \mdelta \, | \, n \in \bbbz
\}$, corresponding to the generators $\{ E_n^{\alpha} \}$.  Affine
algebras also have roots with zero self-intersection, the
$\{ n \, \mdelta \, | \, n \in \bbbz/ \{ 0 \} \}$, corresponding to
$\{ H^i_n \}$.

The imaginary root junction $\mdelta$ has interesting properties.  In
the case of finite algebras on branes, it was not possible to have a
nonzero junction that had vanishing intersection with simple root
junctions and zero $(p,q)$ charge.  However in the affine case,
$\mdelta$ possesses just these traits.  In all the cases examined it
can be presented as a loop of string that encircles the 7-brane
configuration.  Naturally this has no asymptotic charge, and the fact
that it has no Dynkin labels is intuitive since it does not intersect
any of the simple roots, which all lie ``within'' it (see figures in
the next section).  Thus it seems that the condition that a
brane configuration admit an affine algebra can be restated as the
condition that the monodromy matrix admit a nontrivial eigenvector, so
that there exists a $(p,q)$ charge preserved winding around the
branes.

Although the uncharged junctions have degenerate intersection form, the
entire junction lattice is generically nondegenerate for affine Lie algebras.
As a consequence, the affine Cartan matrix cannot be block diagonal within
the junction lattice intersection form, as it was for the finite case.
There is an inevitable ``mixing'' between the uncharged and charged
sectors, which manifests itself in the association of the level $k$
with asymptotic charge, as we now explore.

The level $k$ should remain the same from junction to junction in a
given representation.  In fact, it turns out to be a linear function
of the asymptotic charges, $k=k(p,q)$, which of course also are
constant over a given representation, as the simple roots are
uncharged.  Let us argue that this is reasonable.  For finite Lie
algebras, an arbitrary junction could be characterized completely by
its $r$ Dynkin labels $\{ a_i \}$ and $p$, $q$; the intersection form
block diagonalized into an uncharged ``Lie algebra'' part of
codimension two, and a charged part characterized by $(p,q)$.  For the
affine case there are still two more invariant charges than there are
Dynkin labels; however, the junction $\mdelta$ has vanishing $a_i$ and
$(p,q)$.  There is thus an additional integer $\nt$ which changes when
$\mdelta$ is subtracted from a junction.  We now have $(r+4)$
labels $\{ a_i, p, q, \nt \},\ i = 0 \ldots r$ characterizing a
junction $\mJ$, while the junction lattice is only
$(r+3)$-dimensional.  Since $\nt$ is independent of the others, there
must be a relation among the $\{a_i, p, q \}$.  However, $p$ and $q$
are constant over a representation, while the $\{ a_i \}$ change from
weight to weight.  The only Lie-algebraic quantity that could be
related to the asymptotic charges without putting constraints on the
possible weight vectors is the level $k$, which is also a constant
over a representation.  Thus it is not too surprising that explicit
computation confirms $k=k(p,q)$ in each case.

In summary, the level of the affine weight vector associated to
a junction is determined by the asymptotic charges; equivalently,
fixing the level of a representation puts one constraint on the
asymptotic charges that representation can possess.  After fixing an
affine weight vector, one is still free to choose the other linear
combination of $(p,q)$, as well as $\nt$.

We will employ a basis in which a junction $\mJ$ may be expanded as
\begin{eqnarray}
\mJ = \sum_{i=1}^r a_i \, \mOmega^i + k \, \mOmega^0 + \nt \, \mdelta +
\sigma \, \mSigma \,. \label{j}
\end{eqnarray}
The junctions $\{ \mOmega^i, \mOmega^0 \},\ i = 1 \ldots r$ are dual
to the set $\{ \malpha_i, \mdelta \},\ i = 1 \ldots r$, and thus their
coefficients are the $\{ a_i \}$ and the level $k$.  Of these only
$\mOmega^0$ has asymptotic charge, as it must since it determines $k$.
Both $\mdelta$ and $\mSigma$ are orthogonal to all simple roots.
$\mdelta$ we have already discussed and is uncharged, while $\mSigma$
must have nonzero asymptotic charge; in fact its $(p,q)$ charge is the
same as that of the loop of string realizing $\mdelta$.  (Of course
$\mdelta$ has no {\em asymptotic} charge, it merely carries charge as
it winds around the branes.)  We constrain it to satisfy $(\mdelta,
\mSigma) = 0$.  The coefficient $\sigma$ determines the $(p,q)$
charges not already specified by $k$.

A basis including more familiar fundamental weight junctions $\{
\momega^i \},\ i = 0 \ldots r$ dual to all simple root junctions $\{
\malpha_i \},\ i = 0 \ldots r$ is also possible, but proves less
convenient, as we shall discuss further in the next section.

As the self-intersection $(\mJ, \mJ)$ has been used as a powerful tool
for analyzing the spectra of 3-brane worldvolume theories \cite{DHIZ},
it is useful to consider it for the affine case.  Thus we are
interested in the mutual intersections of $\{ \mOmega^i, \mOmega^0,
\mdelta, \mSigma \}$.  The $\{ \mOmega^i \}$ are orthogonal to
$\mOmega^0$, $\mdelta$ and $\mSigma$, and thus the intersection form
block diagonalizes in this basis.  In the upper block, $(\mOmega^i,
\mOmega^j) = - A^{ij}$ is minus the inverse Cartan matrix of the
finite algebra $g$.  In the lower block, $(\mOmega^0, \mdelta) = -1$
is guaranteed since they are dual.  Recalling $(\mdelta, \mSigma) =
(\mdelta, \mdelta) =0$, we find that $\nt$ will appear only in a
linear term with $k$.  The self-intersection of a junction is then
\begin{eqnarray}
(\mJ, \mJ) &=& - (\overline\lambda, \overline\lambda)_{\mbox{finite}} - 2 \,
\nt \, k + f(p,q) \,, \label{selfintersect} 
\end{eqnarray} 
where $\overline\lambda$ is the horizontal weight vector, and $f(p,q)$
is a quadratic form in $(p,q)$ obtained from reexpressing $k$ and
$\sigma$ in the $(\mSigma, \mSigma)$, $(\mOmega^0, \mOmega^0)$ and
$(\mOmega^0, \mSigma)$ terms, which vary from algebra to algebra.

The grade $n$ of a weight vector is the number of imaginary
roots added to the highest weight, and since $\nt$ is the
coefficient of the imaginary root junction, $n$ will differ from $\nt$
by at most a representation-dependent constant, $\nt_0 = \nt - n$.  We
thus see that the self-intersection includes the affine inner product,
\begin{eqnarray}
(\mJ, \mJ) &=& - (\lambda, \lambda)_{\mbox{affine}} -2 \, \nt_0 \,
k(p,q) + f(p,q) \,,
\end{eqnarray}
and that the remaining contribution is entirely a function of the
$(p,q)$ charges.

There will always be (at least) one brane which, when it is decoupled,
will leave behind the configuration associated to the finite algebra
$g$.  In the basis above, we can require that only $\mdelta$ has
support on this brane.  As a result, the junctions which survive a
decoupling from the affine algebra $\hat{g}$ to the horizontal algebra
$g$ are precisely those with $\nt =0$.  Thus the self-intersection
formula returns to the finite expression, and consequently $f(p,q)$ is
the same quantity that appears in the finite case.

The power of the affine algebra is that junctions must organize
themselves into affine representations.  In specifying a highest
weight representation, one must choose $k$ and a finite highest weight
vector $\overline\lambda_0$.  One linear combination of the asymptotic
charges is now determined by the level.  The other combination of
charges is still free to be chosen (modulo conjugacy constraints, as
we will discuss shortly), as is the integer $\nt_0$, which can be
anything as long as the resulting junction has a good
self-intersection, $(\mJ, \mJ) \geq -2 + \mbox{gcd}(p,q)$.  The grade
of a weight vector in the representation will then be $n = \nt -
\nt_0$.

The affine Weyl group preserves the affine inner product.  If a weight
is in a representation, so is its entire Weyl orbit.  Since the affine
inner product appears as the only contribution of the weight vector to
the self-intersection, entire Weyl orbits will be forbidden or allowed
collectively.  Thus the self-intersection bound will be consistent
with the structure of affine representations.

As is well-known, $k >0$ is necessary for affine highest weight
representations.  For $k <0$ we will instead find lowest weight
representations, which terminate below at an antidominant weight
vector with all Dynkin labels negative, and continue forever in the
direction of increasing grade.  This is in agreement with our
expectations that changing all the signs of the invariant charges (and
thus the weight vectors) of a representation should give us an
identical or conjugate representation, and is consistent with
self-intersection as well, which for $k < 0$ will forbid any affine
weight vector with sufficiently low grade.  Representations with $k=0$
will be infinite in both directions; the adjoint is an example of
this.  In this case self-intersection is independent of the grade
entirely.

Affine Lie algebras inherit unchanged the conjugacy restrictions of
the corresponding finite algebra.  As discussed in \cite{DZ}, when a
Lie algebra has different conjugacy classes, some possible values of
$(p,q)$ for a representation are excluded depending on the conjugacy
class.  This is a result of the fundamental weight junctions
$\mOmega^i$ ($\momega^i$ in \cite{DZ}) not being proper, while
requiring that the junctions in the representation itself are proper.
In general if there are $d$ conjugacy classes, $p$, $q$ or a linear
combination will be restricted to a certain value (mod $d$).  This
remains in the affine case because the ``new'' junction in the
fundamental weight basis is $\mdelta$, which is always proper, as it
is an integral combination of simple roots.

In Table 1, we present a list of {\bf A}-, {\bf B}- and {\bf C}-brane
configurations on which affine Lie algebras arise.  The entire affine
$E_n$ series can be realized for $n \leq 8$, with $\widehat{E_5} =
\widehat{so(10)}$, $\widehat{E_4} = \widehat{su(5)}$, $\widehat{E_3} =
\widehat{su(3)} \times su(2)$, $\widehat{E_2} = \widehat{su(2)} \times
u(1)$, $\widehat{E_1} = \widehat{su(2)}$.  Some can be realized in
more than one way.  Additionally there are a few other affine algebras
of horizontal rank $8$.  Perhaps there are other possibilities.
Definitions for $\mA$-, $\mB$- and $\mC$-branes and their monodromies can be
found in \cite{DZ,ghz}; the associated $[p,q]$ charges are [1,0],
[1,-1] and [1,1] repsectively.

\begin{table}
\begin{center}
\begin{tabular}{|c|c|c|c|} \hline \hline
Brane configuration & Affine algebra & $k=k(p,q)$ & Brane decoupled \\ \hline
& & & \vspace{-4 mm} \\ 
$(\mA^{n-1})\mB\mC\mB\mC$  & $\widehat{E_n}$ & $k=-q$ & $\mb_2$ \\ 
$(\mA^8)\mB\mC\mC$  &   $E_9 \equiv \widehat{E_8}$ & $k=-p+3q$ & $\ma_8$ \\
$(\mA^6)\mB\mC^3$  &  $\widehat{E_7}$ & $k=-p+2q$ & $\mc_1$ \\
$(\mA^4)\mB^2\mC^2$ & $\widehat{so(10)}$ & $k=-p+2q$ & $\mb_1$ \\ 
$(\mA^8)\mB\mC\mB$ & $\widehat{so(16)}$ & $k = -p + q$ & $\mb_2$ \\
$(\mA^8)\mC\mB\mC$ & $\widehat{su(8)} \times su(2)$ & $k=-p + 7p$ & $\mb$ \\
$(\mA^4)\mB(\mA^4)\mC\mC$ & $\widehat{su(8)} \times su(2)$ & $k = -3p + 5q$ & 
$\mb$ \\  \hline
\end{tabular}
\end{center}
\caption{Brane configurations giving rise to affine Lie algebras, including
the relation between the level $k$ and the asymptotic charges $(p,q)$
and the brane which gives the finite algebra upon decoupling.}
\end{table}

Other, more complicated algebras can appear on 7-branes as well.  In
particular, the brane configuration $(\mA^{n-1})\mB\mC\mC$ realizes the $E_n$
series, including not only $E_9 \equiv \widehat{E_8}$ but the hyperbolic
algebra $E_{10}$ and all the rest.  We will not discuss these other
algebras in this paper.

\section{Examples}
\subsection{Affine su(2)}

We now proceed to illustrate these ideas with a discussion of the
brane realization of the simplest affine Lie algebra,
$\widehat{su(2)}$, which appears on the brane configuration
$\mB\mC\mB\mC$.

\begin{figure}
$$\BoxedEPSF{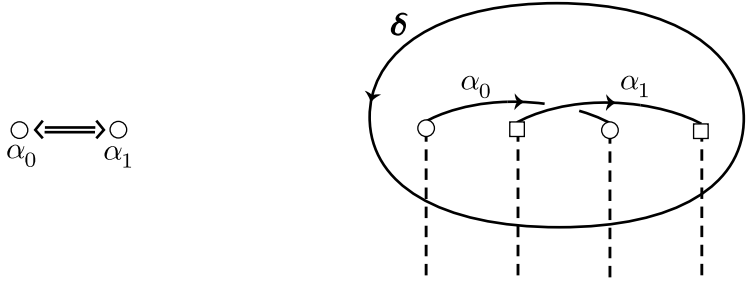}$$
\caption[Affine $\widehat{su(2)}$]{The Dynkin diagram for the
$\widehat{su(2)}$ algebra, where the double line with arrows indicates
an angle of $180^o$ between roots of equal length, and the $\mB\mC\mB\mC$ brane
configuration including simple roots $\malpha_0$, $\malpha_1$ and
the imaginary root $\mdelta$.}
\label{su2}
\end{figure}

 There are four invariant charges, $(Q_B^1, Q_C^1, Q_B^2,
Q_C^2)$.  The intersection form in this basis is
\begin{equation}
\pmatrix{\hskip-3pt -1 & 1 &0 & 1 \cr
                     1 & -1 & -1 & 0 \cr
                     0 & -1 & -1 & 1 \cr
                     1 & 0 & 1 & -1}\,.
\end{equation}
The conditions for uncharged junctions are $Q_B^2 = -Q_B^1$, $Q_C^2 =
- Q_C^1$.  The natural choice for a basis for the uncharged junctions
is the set of simple roots
\begin{eqnarray}
\malpha_0 = \mb_1 - \mb_2 \,, \quad
\malpha_1 = \mc_1 - \mc_2 \,,
\end{eqnarray}
having the intersection form
\begin{equation}
\pmatrix{\hskip-2pt -2 & 2 \cr 2 & -2}\,
\end{equation}
which is indeed the negative of the $\widehat{su(2)}$ Cartan matrix.

The Coxeter labels for $su(n)$ algebras are all $1$, and hence $\mdelta =
\malpha_0 + \malpha_1$.  Indeed we can confirm that $(\mdelta, \malpha_i) =0$.
The monodromy matrix for these branes is 
\begin{equation}
\pmatrix{\hskip-3pt 1 & 8 \cr 0 & 1}\,
\end{equation}
which admits the eigenvector $(1,0)$; $\mdelta$ can be presented as a $(1,0)$
string winding around the $\mB\mC\mB\mC$ configuration.

Taking the intersection $(\mdelta,\mJ)$ for an arbitrary junction
$\mJ$, we have $a_0 + a_1 = k = Q_B^1 - Q_C^1 + Q_B^2 - Q_C^2 = -q$,
the desired relation between the level and the asymptotic charges.
Specifying the affine weight vector by $a_0$, $a_1$ thus
fixes $q$ as well, leaving $p$ and $\nt$ to be determined.

One would like to expand an arbitrary junction in a basis such that
the Dynkin labels, charges and $\nt$ are the expansion coefficients.
Before we discuss our preferred basis $\{ \mOmega^i, \mdelta, \mSigma
\}$, let us explore the more natural place to start, namely junctions
$\{ \momega^0, \momega^1 \}$, dual to $\{ \malpha_0, \malpha_i \}$ and
satisfying the usual inner products of fundamental weights:
$(\momega^1, \momega^1) = -1/2$, $(\momega^0, \momega^0) = (\momega^0,
\momega^1) = 0$.  One would then add to these $\mdelta$ and a charged
junction $\mSigma$, both orthogonal to simple roots:
\begin{eqnarray}
\mJ = a_0 \, \momega^0 + a_1 \, \momega^1 + \nt \, \mdelta + \sigma \, \mSigma 
\,.
\end{eqnarray}
Notice that since the level is $k = -q$, $\sigma$ is a combination of
asymptotic charges including $p$.

There are many choices of invariant charges for such junctions, one of
which is
\begin{eqnarray}
\momega^0 &=& (-\fracs14,-\fracs34,\fracs14,-\fracs14) \,, \nonumber \\  
\momega^1 &=& (-\fracs14,-\fracs14,\fracs14,-\fracs34) \,, \\
\mdelta &=&  (1,1,-1,-1) \,, \nonumber \\
\mSigma &=&  (\fracs14,0,\fracs14,\fracs12)   \,.  \nonumber
\end{eqnarray}
There are disadvantages to this presentation.  In particular, $\nt =
\fracs34 Q_B^1 - \fracs14 (Q_C^1 + Q_B^2 + Q_C^2)$, which generically
is not an integer, a situation which is not disastrous but is
inconvenient.  Additionally, we are interested in the condition $Q_B^2
= 0$ for a junction surviving the decoupling of a $\mB$-brane to the
non-affine $su(2)$, $\mB\mC\mC$ configuration, which in this basis has the
arcane form $\nt = \fracs14 (p + 2k)$.  No other realization of the
$\{ \momega^i \}$ improves the situation.  In the $\{ \mOmega^i \}$ basis,
however, $\nt$ is integral and the condition for decoupling is simple.

We take the junctions $\{ \mOmega^0, \mOmega^1 \}$ to be dual to
$\{ \mdelta, \malpha_1 \}$.  As in (\ref{j}), a junction is given by
\begin{eqnarray}
\mJ &=& a_1 \, \mOmega^1 + k \, \mOmega^0 + \nt \, \mdelta + p \, \mSigma \,,
\end{eqnarray}
where $k$ now appears because $(\mOmega^0, \mdelta) = -1$, and
$\sigma = p$ in this basis.  The basis junctions are
\begin{eqnarray}
\mOmega^0 &=& (0,\fracs12,0,-\fracs12) \,, \nonumber \\  
\mOmega^1 &=& (\fracs12,-\fracs14,0,-\fracs14) \,, \\
\mdelta &=&  (1,1,-1,-1) \,, \nonumber \\
\mSigma &=&  (\fracs12,\fracs14,0,\fracs14)   \,.  \nonumber
\end{eqnarray}
We now have the simple relation $\nt = - Q_B^2$, which is always an integer for
proper junctions, and implies $\mJ$ survives decoupling precisely when
$\nt(\mJ) = 0$.  This is very convenient, as the integer $\nt$ measures 
directly the affine support of a given junction.

Self-intersection in this basis is
\begin{eqnarray}
(\mJ, \mJ) &=& - \fracs12 (a_1)^2 - \fracs78 k^2 - 2 \nt k - \fracs14 k p +
\fracs18 p^2 \,, \label{jsu2} \\
&=& - \fracs12 (a_1)^2 - 2 \nt k + \fracs18 (p^2 + 2pq - 7 q^2) \,, \nonumber
\end{eqnarray}
reproducing the form (\ref{selfintersect}).  

To find the junction realization of a given representation, we specify
the representation by the level $k$ and the Dynkin label $a_1$ of the
highest weight.  $q = -k$ is now determined.  There are two conjugacy
classes in $su(2)$, given by $a_1 \; (\mbox{mod}\ 2)$.  Requiring an
arbitrary junction to be proper results in the condition
\begin{eqnarray}
\fracs12(p+q) = a_1 \; (\mbox{mod}\ 2) \,,
\end{eqnarray}
where junctions with $p+q$ odd cannot be realized at all.  Lastly one
may pick any $\nt_0$ for the highest weight as long as $(\mJ, \mJ)
\geq -2 + \mbox{gcd}(p,q)$.  Other weights are found by subtracting
$\malpha_i$, and $n = \nt - \nt_0$.

Consider for example the representations at $k=1$.  The level fixes $q
= -1$.  We can have $(a_0, a_1)$ either $(1,0)$ or $(0,1)$ for the
highest weight.  For $(1,0)$, $p-1 = 0 \; (\mbox{mod}\ 4)$.  We can
for example choose $p=1$, in which case $f(p=1,q=-1) = -1$. Also
$\mbox{gcd}(p=1,q=-1) =1$, and so finally we find $(\mJ_0, \mJ_0) = -2
\nt_0 -1$ and $\nt_0 \leq 0$.  We could choose $\nt_0 = 0$, in which
case $n = \nt$ and the only junction in the entire representation to
survive decoupling to finite $su(2)$ is the $su(2)$ singlet at $\nt =
0$, which has $Q_B^1 = 1$, others zero.  For other choices of $\nt_0$,
no junctions satisfy $\nt=0$ and the entire representation fails to
survive decoupling.

A representation with $k=1$ and highest weight $(0,1)$ requires $p - 1 = 2 \;
(\mbox{mod}\ 4)$.  Let us this time pick $p=-5$, which gives $f(p=-5,q=-1) =
7/2$ and $\nt \leq 2$.  Choosing $\nt_0 = 2$ means that the junctions
which survive decoupling exist at $n=-2$ in this representation; the
finite $su(2)$ content is a spin-3/2 and a spin-1/2 \cite{slansky}.
Other choices of $\nt_0$ produce other finite pieces upon decoupling.

\subsection{Affine E8}

\begin{figure}
$$\BoxedEPSF{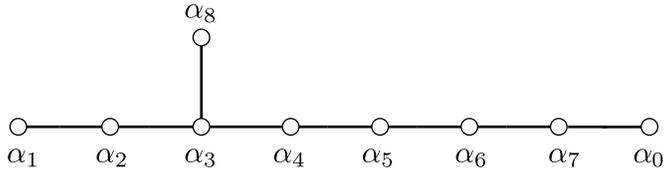}$$
\caption[Affine $\widehat{E_8}$]{The Dynkin diagram for the
$\widehat{E_8}$ algebra.}
\label{e8}
\end{figure}

Having seen how the minimal affine algebra $\widehat{su(2)}$ is realized
on branes, we now turn to explore as a final example $\widehat{E_8}$.
The Dynkin diagram is displayed in Figure~\ref{e8}.

The finite $E_8$ algebra can be realized on the brane configuration
$(\mA^7)\mB\mC\mC$ \cite{GZ}, and we follow the conventions for simple
roots of \cite{DZ}.  The affine Lie algebra $\widehat{E_8}$ can be
realized in more than one way.  The finite $su(2)$ and its affine
counterpart $\widehat{su(2)}$ are part of a series covering all the
exceptional algebras where the finite $E_n$ algebra is realized on
$(\mA^{n-1})\mB\mC\mC$ and $\widehat{E_n}$ is realized on
$(\mA^{n-1})\mB\mC\mB\mC$, with the quadratic form $f(p,q) =
\frac{1}{9-n} (p^2 - (n-3) pq + (2n-9) q^2)$.  Hence $\widehat{E_8}$
appears on $(\mA^7)\mB\mC\mB\mC$.  Additionally, if one tries to
extend the finite series beyond $E_8$, one gets to $E_9$ which is
simply $\widehat{E_8}$ again; the algebra thus also appears on
$(\mA^8)\mB\mC\mC$.  In all cases the asymptotic charge quadratic form
is $f(p,q) = p^2 - 5pq + 7q^2$.

\begin{figure}
$$\BoxedEPSF{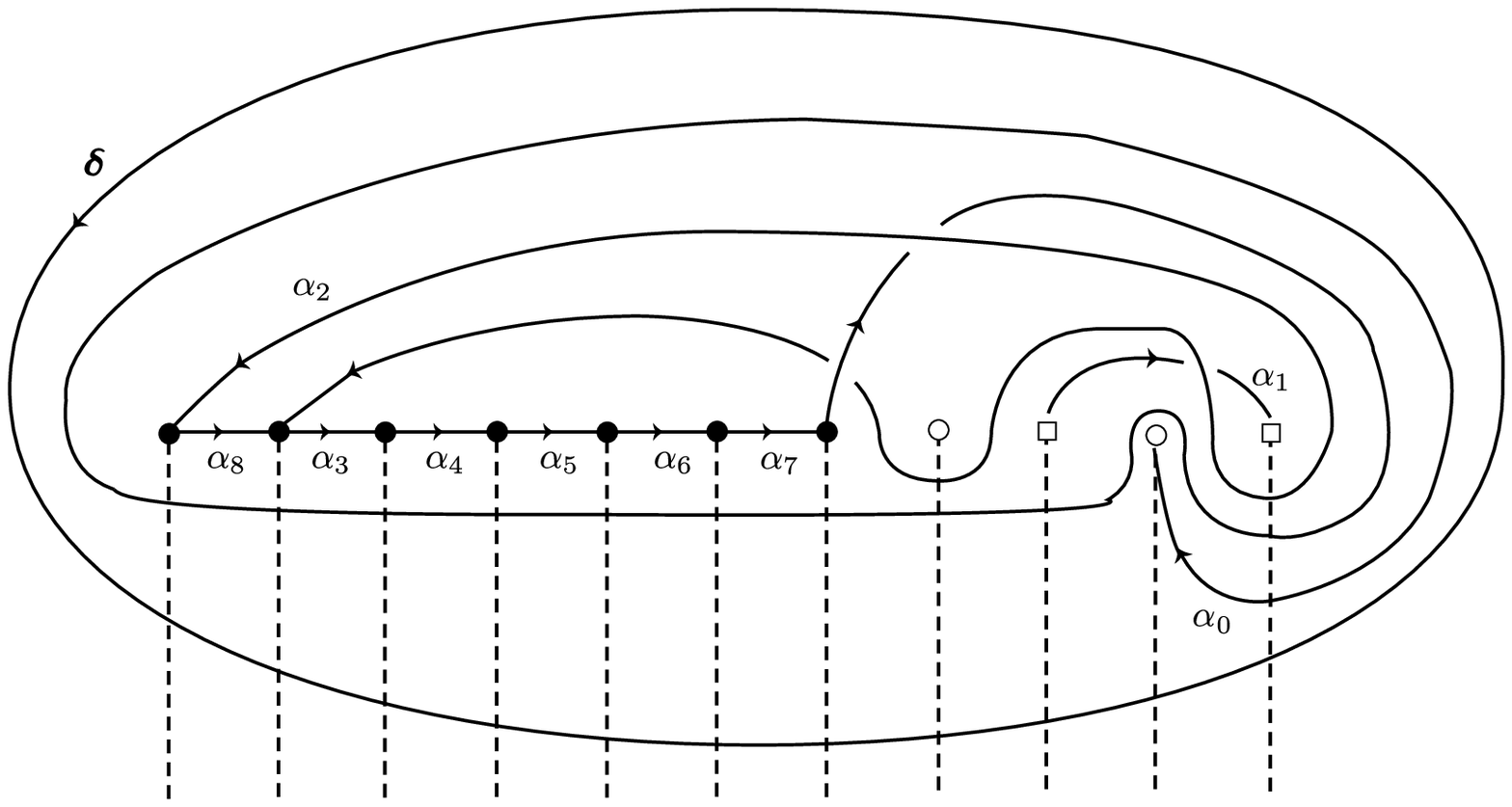 scaled 750}$$
\caption[First configuration for $\widehat{E_8}$]{The $\widehat{E_8}$ algebra
realized on the brane configuration $(\mA^7)\mB\mC\mB\mC$, with the simple roots
$\malpha_i$ and imaginary root $\mdelta$ indicated.}
\label{bcbc}
\end{figure}

While $f(p,q)$ coincides for both realizations of $\widehat{E_8}$,
since both must reduce to the same $E_8$ upon the appropriate
decoupling, the appearance of $\mdelta$ is different.  For the entire
$(\mA^{n-1})\mB\mC\mB\mC$ series, $\mdelta = \mb_1 + \mc_1 - \mb_2 -
\mc_2$, which can be realized as a $(1,0)$ string winding around the
entire configuration; since it is insensitive to $A$-brane monodromies
the appearance of new $A$-branes does not disturb it.  As a result $k
= -q$ for all these configurations, including the
$(\mA^7)\mB\mC\mB\mC$ $\widehat{E_8}$.  $\{ \malpha_i\}, i = 1 \ldots
8$ are the same as the finite case, and $\malpha_0$ is then
\begin{eqnarray}
\malpha_0 = \mdelta - \mtheta = - \sum_{i=1}^7 \ma_i - \ma_7 + 5 \mb_1 + 3
\mc_1 - \mb_2 + \mc_2 \,, 
\end{eqnarray}
and the $\widehat{E_8}$ Cartan matrix is reproduced.  The simple root
junctions and the imaginary root $\mdelta$  on this brane configuration
are shown realized as simple strings in Figure~\ref{bcbc}.

\begin{figure}
$$\BoxedEPSF{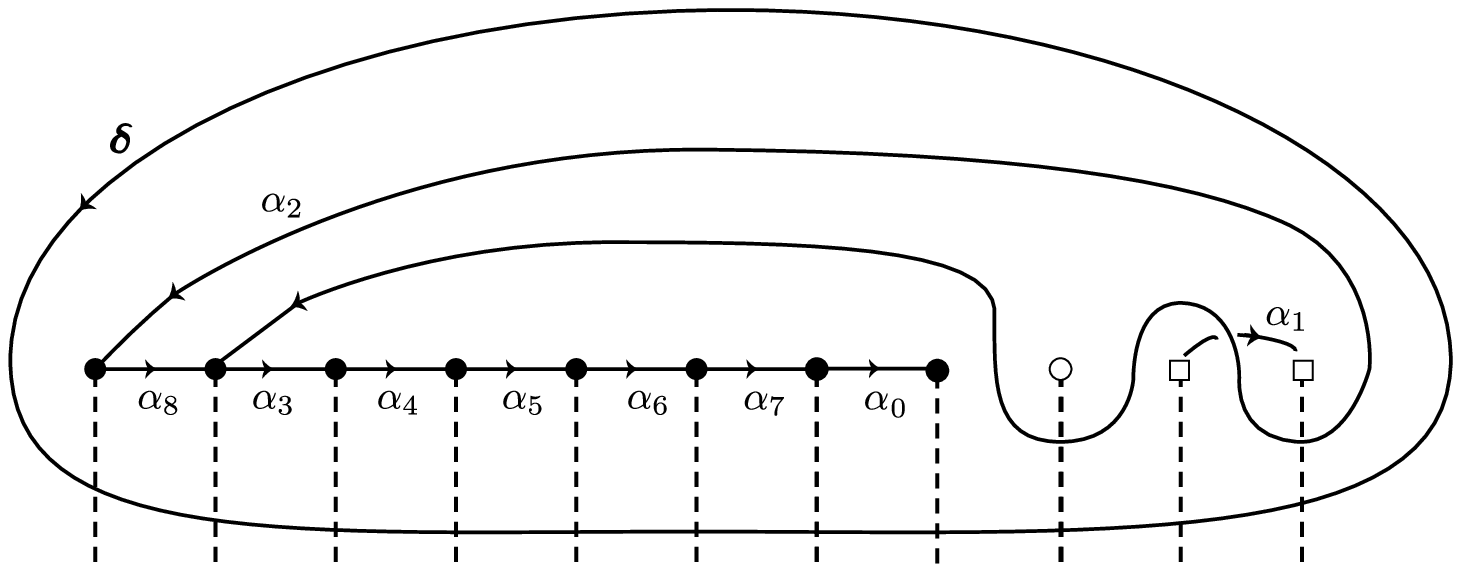 scaled 750}$$
\caption[Second configuration for $\widehat{E_8}$]{The $\widehat{E_8}$
algebra realized on the brane configuration $(\mA^8)\mB\mC\mC$, with
the simple roots $\malpha_i$ and imaginary root $\mdelta$ indicated.}
\label{bcc}
\end{figure}

On the other hand, the monodromy of the $(\mA^8)\mB\mC\mC$ branes
admits the eigenvector $(3,1)$, and there is a $\mdelta$ junction
realized by a $(3,1)$ string winding around the branes.  This junction
coincides with that obtained by defining $\malpha_0 = \ma_7 - \ma_8$
and using
\begin{eqnarray}
\mdelta = \malpha_0 + \mtheta = - \sum_{i=1}^8 \ma_i + 4 \mb + 2 \mc_1 + 2
\mc_2 \,.
\end{eqnarray}
Calculating $k = - (\mJ, \mdelta)$ for an arbitrary $\mJ$, we arrive at
$k = -p + 3q$.  The junctions for this configuration are displayed in
Figure~\ref{bcc}.

In both cases, we can expand a junction in the decoupling basis
\begin{eqnarray}
\mJ &=& \sum_{i=1}^8 a_i \, \mOmega^i + k \, \mOmega^0 + \nt \, \mdelta + \sigma
\mSigma \,,
\end{eqnarray}
and $(\mJ, \mJ)$ is given by (\ref{selfintersect}) with $f(p,q)$ given
above.  In both cases $\nt = 0$ gives the condition for junctions to
survive decoupling; of course, $k$ is different for each.  For
$(\mA^7)\mB\mC\mB\mC$ we have $\sigma = p$, while for
$(\mA^8)\mB\mC\mC$ we choose $\sigma = q$.

$E_8$ has no conjugacy classes and thus neither does $\widehat{E_8}$.

Let us explore an example of a representation.  Consider junctions of
charge $(-1,-1)$ in the $(\mA^7)\mB\mC\mB\mC$ presentation.  The
charge quadratic form is $f(p=q=-1)=3$.  The level is fixed at $k=1$.
There is only one affine representation at this level, with highest
weight $a_0 =1, a_i= 0, i \neq 0$.  For this weight $(\mJ, \mJ) = -2
\, \nt_0 + 3$, implying $\nt_0 \leq 2$.

In \cite{DHIZ}, worldvolume theories for D3-branes in the vicinity of
7-brane configurations were considered.
It was seen that the ${\bf 3875}$ representation must
be in the spectrum of the finite $E_8$ algebra theory, as a result of
consistency with the known $D_4$ spectrum, which is just Seiberg-Witten
theory with $N_f = 4$ flavors.  In the affine case, choose $\nt_0 = 2$.
The junctions that survive decoupling to the finite case are then at
grade $n=-2$ of this representation, which has $E_8$ content
\cite{slansky}:
\begin{eqnarray}
{\bf 3875} \oplus {\bf 248} \oplus {\bf 1} \,.
\end{eqnarray}
We then know that the ${\bf 248}$ and ${\bf 1}$ must be in the $E_8$
spectrum as well, since the ${\bf 3875}$ which is known to be present
must lift to some complete affine representation, and as this is the
only possibility must be accompanied by the ${\bf 248}$ and ${\bf 1}$.
Other choices of $\nt_0$ lead to other representations, with smaller
or no surviving horizontal content.  The ${\bf 3875}$ does not appear
and so makes no statement about the existence of these
representations.

A similar process occurs for the $(\mA^8)\mB\mC\mC$ theory, where the
$(-1,-1)$ charges will be at level $k = -2$.  There the ${\bf 3875}$
will lift to the $n=0$ grade of a lowest weight representation, and it
is the only horizontal representation at that grade.

Before concluding, we note that one could inquire about the brane
configuration $(\mA^8)\mB\mC\mB\mC$, which would be ``affine $E_9$''.  This
algebra is the combination of two $so(8)$ singularities as studied by
Imamura \cite{imamura} and cannot be realized as a singularity on K3
without destroying the triviality of the canonical bundle.  It can be
thought of as finite $E_8$ with two different imaginary root
junctions, $\mdelta_1$ and $\mdelta_2$, which have vanishing
intersection with the $E_8$ simple roots, themselves and each other.
Both can arise because the monodromy matrix is unity and thus any
$(p,q)$ string can wind around the configuration.  The uncharged
junction sublattice has dimension 10, and so it cannot be any affine
algebra, which have rank $\leq 9$; but the Cartan matrix is
degenerate, so neither is it hyperbolic (in particular it is not
$E_{10}$).  It is not clear what kind of Lie algebra this is.

\section{Conclusions}

We have considered the realization of affine Lie algebras as string
junctions on configurations of 7-branes.  We have discussed how Lie
algebras arise from the intersection form of the holomorphic curves
associated to the junctions in the M/F-theory picture, when this form
contains the appropriate Cartan matrix.  When an affine Cartan matrix
is realized, the junction associated to the imaginary root manifests
itself as a loop of string whose charges are a nontrivial eigenvector
of the branes' monodromy matrix.  The level of the affine algebra
is equal to a linear combination of the asymptotic charges, and thus
is assured to be a constant in a given representation.  

The intersection form includes the full affine inner product.  Unlike
the inner product of finite Lie algebras, the affine version is of
indefinite signature.  It arises naturally as part of the junction
intersection form, which has indefinite signature for both finite and
affine cases.  In the finite case the positive and negative
eigenvalues are segregated into ``Lie algebra'' and ``asymptotic
charge'' blocks.  In the affine case however, the affine Cartan matrix
is degenerate and so cannot be block diagonal within the intersection
form; the off-diagonal elements mix the positive and negative
eigenvalues and this results in the appearance of the affine inner
product, as well as combining the Lie algebra and charge sectors so as
to fix the relation between $k$ and $(p,q)$.

The brane configuration associated to an affine algebra cannot
coalesce to a single point on K3, but junctions must nonetheless fill
out representations of the affine algebra.  These representations are
infinite-dimensional, but only a finite number of junctions can be
massless.  It has been only partially known which representations of
finite algebras actually appear as BPS states in 3-brane worldvolume
field theories; the structure of the affine representations and the
known existence of a few finite representations requires the presence
of many more.

We have laid the Lie-algebraic groundwork for understanding affine
algebras on 7-branes.  Many applications remain to be examined.  A
more complete study of which representations are required in the
spectra of the 4D, $E_n$ field theories still remains to be done.
Assuming these theories to be the same as those obtained by
compactifying the 6D non-critical string on a torus, a fuller
geometrical investigation of the duality between these two
configurations would be interesting.  This could conceivably involve
the $(\mA^8)\mB\mC\mB\mC$ brane configuration which characterizes a ${\cal B}_9$
del Pezzo surface, or ``$\fracs12 K3$''.  The non-critical string
theory and its compactifications are still poorly understood, but the
prospects for investigating them are becoming brighter.

\subsection*{Acknowledgments}

I am grateful to Tam\'as Hauer, Amer Iqbal, Cumrun Vafa and Barton
Zwiebach for illuminating discussions and helpful suggestions, and to
Kristin Burgess and Alexandria Ware for editing the manuscript.  This
work was supported by the U.S.\ Department of Energy under contract
\#DE-FC02-94ER40818.

\end{document}